\documentclass[journal]{IEEEtran}

\usepackage{cite}

\ifCLASSINFOpdf
\usepackage[pdftex]{graphicx}
\usepackage{amsmath}
\usepackage{amssymb}
\usepackage{balance}
\newcommand{\subcaption}[1]{\begin{center}#1\end{center}}
\usepackage{threeparttable}
\else
\fi

\usepackage{color}

\begin{document}
%
\title{Self-heating in SIS Mixers: Experimental
Evidence and Theoretical Modeling}
%
%
%

\author{Wenlei Shan, \IEEEmembership{Member,~IEEE}
and~Shohei~Ezaki,~\IEEEmembership{}
\thanks{This work was partly supported by the Japan Society for the Promotion of Science (JSPS) KAKENHI under Grant Number 21H01134 and 22H04955}
\thanks{Wenlei Shan, and Shohei Ezaki are with National Astronomical Observatory of Japan(NAOJ), Osawa 2-21-1, Mitaka, 181-8588, Tokyo, Japan (e-mails: wenlei.shan@nao.ac.jp;shohei.ezaki@nao.ac.jp;).  Wenlei Shan is also with the Graduate University for Advanced Studies (SOKENDAI).}
}
%
%

\markboth{Submission ID: XXX}%
{XXX}
%


\maketitle

\begin{abstract}
 This work investigates the relationship between self-heating and the characteristic features observed in the current–voltage characteristics (IVCs) of superconductor–insulator–superconductor (SIS) junctions. Finite-element analysis is employed to evaluate the steady-state temperature distribution around SIS junctions, explicitly accounting for the temperature dependence of the thermal conductivities of the constituent materials. This approach enables flexible estimation of self-heating under various practical conditions, such as different substrate materials, interfacial thermal resistances, and geometric layouts. A heating coefficient is extracted from the simulations and used as an input parameter for IVC modeling. Incorporating self-heating through temperature-dependent gap energy and quasiparticle broadening, the simulated IVCs reproduce bending features near the energy gap that agree with measured characteristics. Furthermore, when a weak link is present near an SIS junction, its critical current can be significantly reduced by junction heating, producing unexpected bends at the linear branch of measured IVCs. Conversely, such bends may serve as indicators that the junction temperature approaches the superconducting transition temperature.
\end{abstract}

\begin{IEEEkeywords}
Superconductor–Insulator–Superconductor junction, self-heating, finite-element analysis, current–voltage characteristics.
\end{IEEEkeywords}

%
\IEEEpeerreviewmaketitle

\section{Introduction}
%
%
%
%
\IEEEPARstart{S}uperconductor–insulator–superconductor (SIS) tunnel junctions are widely employed in digital circuits, quantum computing, and weak-signal detectors. In the present context, they are used as heterodyne mixers for detecting millimeter- and sub-millimeter-wave signals in radio astronomical observations, owing to their quantum-limited intrinsic noise. To achieve broadband response, SIS mixers require increased current density to suppress the capacitive component of the junction impedance, which is the dominant frequency-dependent factor. Typical current densities are 10~kA/cm$^{2}$ for AlO$_x$ barriers \cite{kroug2009sis} and 20~kA/cm$^{2}$  for AlN barriers \cite{Bumble2001Fabrication}. Such high current densities raise concerns about self-heating under dc bias and local oscillator (LO) pumping.

For a rough estimate, a 2-µm junction at 10~kA/cm$^{2}$ has a normal resistance of approximately 5~$\Omega$. Under optimum pumping conditions, the LO power, proportional to the square of the LO frequency, is about 0.1~µW at 140~GHz and 1.4~µW at 500~GHz. The dc power in both cases is about 0.3~µW. Although these power levels appear small, as will be shown later, significant junction heating can still occur under certain conditions.

\subsection{Theory Revisit}
Self-heating in SIS mixers was first analyzed theoretically in~\cite{dieleman1996direct} using the thermal diffusion equation, a method also applied in related works~\cite{leone2000geometric, leone2002hot, leone2001electron,kittara2009measurement}. With simplified geometry and temperature-independent thermal conductivity, the model yields an analytical solution in which the exponential decay of temperature around a heated junction is characterized by the healing length
\begin{equation}
\eta = \sqrt{\kappa d / Y_K},
\end{equation}
where $\kappa$ is the thermal conductivity of Nb, $d$ is the thickness of the wiring layer, and $Y_K$ is the interfacial (Kapitza) thermal resistance between the base electrode and substrate.

While this approach captures the essential physics, it leaves several key issues unresolved. First, since thermal conductivities vary strongly with temperature in the relevant regime, the resulting temperature distribution induces a conductivity distribution, requiring a numerical rather than analytical solution. Second, when a junction is biased along the linear branch of the IVC, the temperature rise may not be negligible compared to the bath temperature, leading to significantly enhanced thermal conductivity and making further heating more difficult. This introduces uncertainty in comparing theoretical photon-step widths with experimental results above the gap voltage~\cite{dieleman1996direct}.

Other discrepancies between the simplified theory and experiment include:
\begin{itemize}
\item The prediction that increasing the wiring-layer thickness should improve cooling, which is not observed in our measurements.
\item The prediction of substrate independence due to large Kapitza resistance, while our experiments reveal clear substrate-dependent differences.
\end{itemize}
These limitations motivate a numerical solution of the diffusion equation that accounts for the temperature dependence of thermal conductivities.

\subsection{Relation Between Broken IVCs and Self-Heating}
\begin{figure}[tb]
\centering
\includegraphics[width=3.2in,clip]{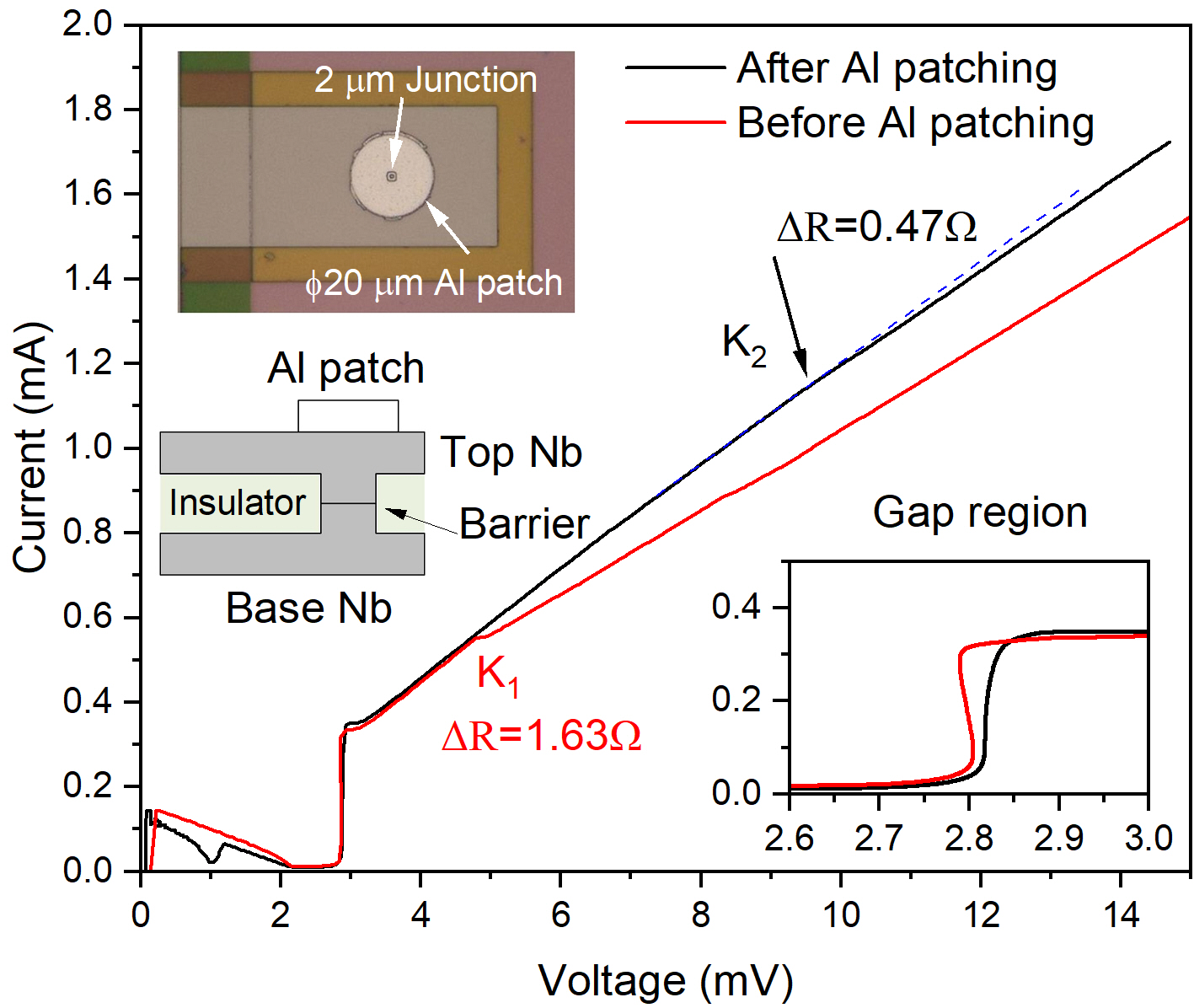}
\caption{Example showing an IVC with a clear bend at $K_1$ on the linear branch. After adding an Al patch on top of the wiring layer, the bend moves upward to $K_2$, and the back-bending feature at the gap voltage becomes less significant. The inset shows an enlarged view of the IVCs near the gap voltage. Also included are a microscopic image of the device and a schematic cartoon illustrating the device's cross section and the position of the Al patch. }
\label{FigAlPatchMeasurement}
\end{figure}

Unexpected bends have been observed along the linear branches of SIS mixer IVCs, as shown in Fig.~\ref{FigAlPatchMeasurement}. Hereafter, such IVCs are referred to as \emph{broken IVCs}. These bends, often occur in the voltage range from 1.5 to 3~$V_{\text{gap}}$. We observed that adding an aluminum cap patterned above the Nb wiring layer, with a diameter of 20~µm as shown in the inset of Fig.~\ref{FigAlPatchMeasurement}, shifts the bend to higher voltage and simultaneously reduces back-bending near the gap. This suggests a strong connection to junction self-heating. The bend can plausibly be attributed to a weak link near the junction: when the junction is heated toward the weak link’s $T_c$, it switches to the normal state, adding resistance to the linear branch. Interestingly, replacing the cap with a 500-nm-thick Nb patch produced no appreciable effect, indicating that wiring-layer thickness does not effectively improve junction cooling as predicted in~\cite{dieleman1996direct}.

\subsection{Junction Fabrication Process}
The SIS junctions exhibiting broken IVCs were fabricated using a machine-aligned junction etching process. A Nb/Al/AlO$_x$/Al/Nb multilayer was deposited \emph{in situ} and patterned by ICP-RIE. The Nb base and top electrodes were $\sim$200~nm and $\sim$100~nm thick, respectively. Symmetric Al layers of $\sim$5~nm each were included. The junction area was defined by etching through the Al layers into the Nb base electrode. A 300-nm-thick SiO$_2$ layer was deposited by PECVD, followed by via-hole definition and etching. Finally, the wiring layer was deposited and patterned. Details are given in~\cite{ezaki2020fabrication}. Some older devices fabricated using a self-aligned lift-off process~\cite{noguchi1994fabrication} did not show broken IVCs, suggesting that weak links are likely introduced at the via-holes.

\subsection{Organization of This Article}
The remainder of this paper is organized as follows. Section~II introduces the finite-element analysis and compares simulations with measurements. Section~III presents IVC modeling incorporating self-heating. Section~IV discusses the origin of weak links, their connection with junction heating, and their sensitivity to annealing. Section~V considers the impact of self-heating and weak links on SIS mixer performance. Section~VI concludes the paper.

\section{Finite-Element Simulation of Junction Self-Heating}

\subsection{Finite-Element Analysis Method}

\subsubsection{Temperature-Dependent Thermal Conductivity}

\begin{figure}[tb]
\centering
\includegraphics[width=3.2in,clip]{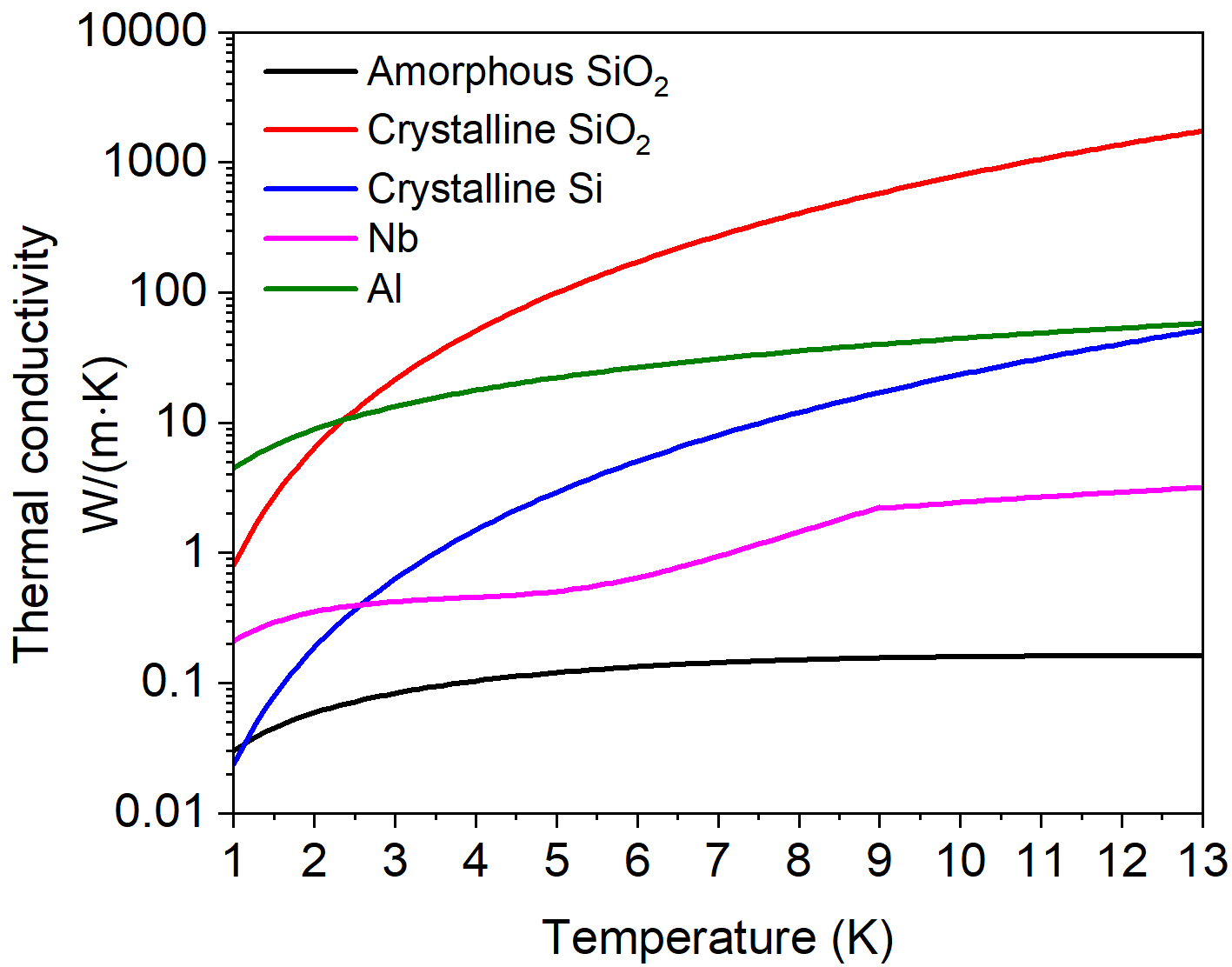}
\caption{Temperature-dependent thermal conductivities of the materials used in the simulations. }
\label{FigConductLog}
\end{figure}

The thermal conductivities of materials used in the simulations are shown in Fig.~\ref{FigConductLog}. Amorphous SiO$_2$ appears as substrate, insulator, or as the surface layer of thermally oxidized silicon substrates. Crystalline SiO$_2$ and crystalline silicon are used as substrates, while Nb and Al serve as electrodes and cooling patches.

The conductivity of amorphous SiO$_2$ is fitted from measurements reported by Simon~\cite{simon1994cryogenic}. For crystalline SiO$_2$ at 4~K, the conductivity is estimated using the kinetic theory $\kappa(4\,\text{K}) = C v l/3 = 51$~W/(m·K), with $C(4\,\text{K}) = 97$~J/(m$^3$K), $v = 5800$~m/s (sound velocity), and $l = 0.3$~mm (mean free path limited by substrate thickness)~\cite{asheghi1998temperature}. Its temperature dependence is approximated by $\kappa(T) = \kappa(4\,\text{K})(T/4)^3$. For crystalline silicon, with $C(4\,\text{K}) = 75.5$~J/(m$^3$K), $v = 9620$~m/s, and $l = 6$~µm (SOI device-layer thickness), the estimated thermal conductivity at 4~K is $\kappa(4\,\text{K}) = 1.5$~W/(m·K), also following a $T^3$ law.

For Nb films, all available measurements correspond to high-purity bulk Nb, and therefore are not directly usable. We estimate the normal-state conductivity using the Wiedemann–Franz law, $\kappa(T) = L T \sigma \approx 1.76$~W/(m·K), with the Lorenz number $L = 2.44 \times 10^{-8}~\rm{W\Omega/K^2}$ and residual resistivity $\sigma = 12.5~\rm{\mu\Omega\cdot cm}$. Below $T_c$, conductivity is scaled by the ratio measured in~\cite{wasim1969thermal}. Aluminum is assumed to remain normal in our temperature range, with conductivity again estimated via the Wiedemann–Franz law.

\subsubsection{Interfacial thermal resistance}

The interfacial thermal resistance between the base Nb electrode and the substrate is one of the dominant factors in determining the extent of junction self-heating~\cite{dieleman1996direct}. Despite its importance, reliable cryogenic measurements of this parameter are scarce. For Nb/silica interfaces, Kapitza conductances in the range of $5 \times 10^{3}$ to $1.7 \times 10^{4}$~W/(K·m$^{2}$) have been reported near $T_c$~\cite{abaloszewa2023thermomagnetic}. However, to the best of our knowledge, no measurement data exist for interfaces between Nb and crystalline substrates such as crystalline SiO$_2$ or crystalline silicon.

Little’s theoretical analysis~\cite{little1959transport} provides useful qualitative guidance. He pointed out that if the two solids have similar acoustic impedances, the interfacial thermal resistance can be approximated by the thermal resistance of each material across a thickness equal to its mean free path. While this model cannot be directly applied in practical calculations due to the difficulty of estimating mean free paths, it strongly suggests that the Kapitza resistance is highly sensitive to the crystallinity and microstructure of the materials forming the interface. Specifically, crystalline substrates are expected to exhibit much higher interfacial conductance than amorphous ones.

Because of this uncertainty and the lack of direct measurement data, in our simulations the Kapitza resistance between base Nb and crystalline substrates was treated as a free parameter. Its value was adjusted to achieve consistency between simulation results and experimental observations of junction heating.

\subsubsection{Simulation Tool and Models}

We used the commercial finite-element software \emph{AMaze}~\cite{AMazeSoftware} to solve the heat equation with tabulated, temperature-dependent conductivities. An iterative scheme adjusts conductivity according to local temperature.

\begin{figure}[tb]
\centering
\includegraphics[width=3.2in,clip]{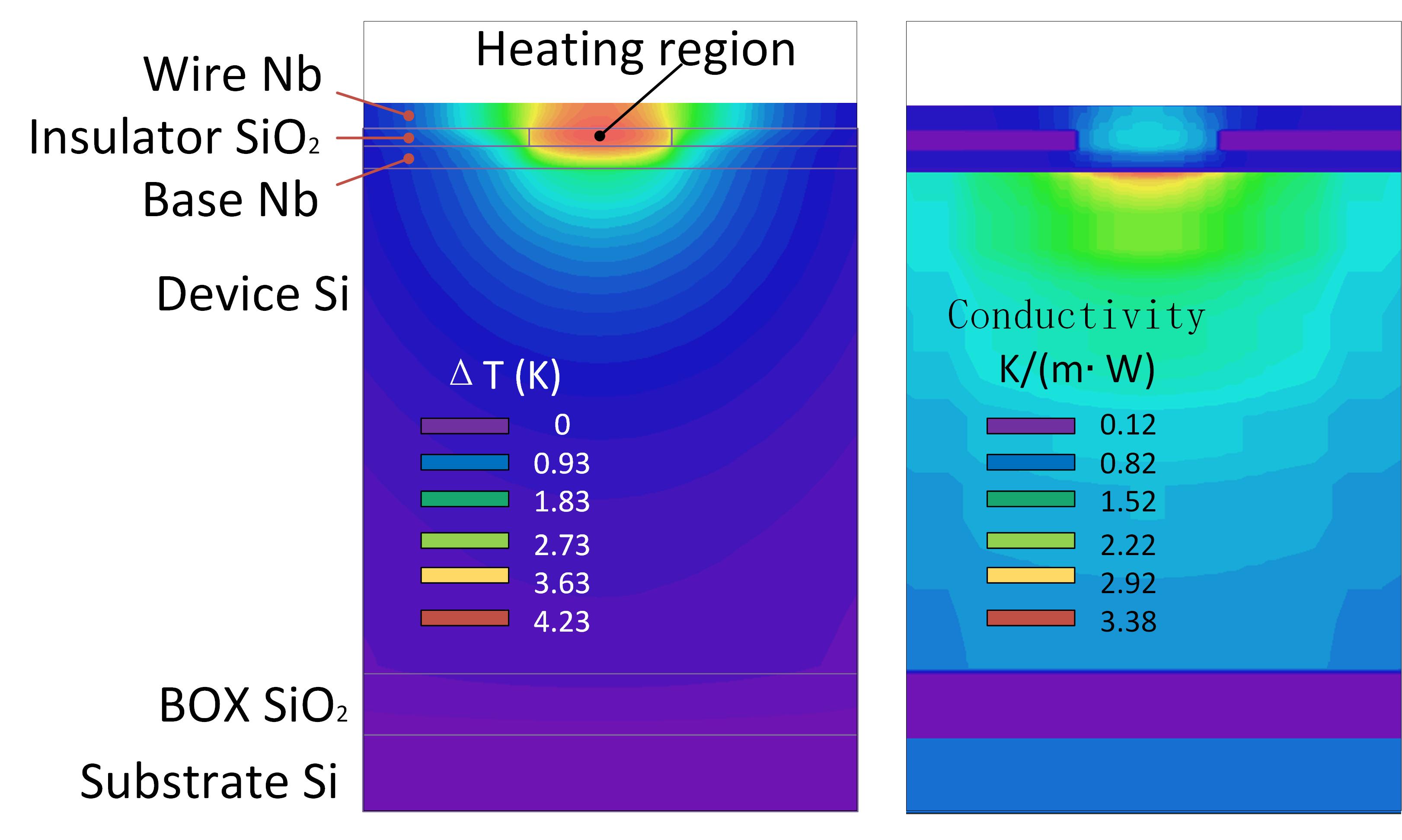}
\caption{Finite-element simulation model of a 2-µm junction on an SOI wafer. Left: temperature rise distribution. Right: thermal conductivity distribution. The artificial interfacial layer used to represent Kapitza resistance is included in the model.}
\label{FigModel}
\end{figure}

Fig.~\ref{FigModel} shows an example temperature map (left) and conductivity distribution (right) around a 2-µm junction  on an SOI wafer. In the model, the base Nb, insulator, and wiring layers are 200~nm, 300~nm, and 400~nm thick, respectively. Heating is applied as volumetric power density in a 300-nm-thick cylinder at the junction. Since the solver does not natively support interfacial resistance, a 100-nm artificial layer with assigned conductivity $k$ was inserted, equivalent to the desired Kapitza resistance $\kappa/t$. Because this layer is thin, the increase in the lateral thermal conductance can be safely neglected.

The SOI substrate model includes a 6-µm Si device layer and a 1-µm buried SiO$_2$ layer. Simulation space extends $\pm15$~µm, much larger than the thermal healing length. Only the central $\pm3$~µm region is displayed in Fig.~\ref{FigModel}.

\subsection{Simulation Results and Comparison With Experiment}

\subsubsection{Substrate Dependence}

A particularly important observation, which contradicts the prediction in~\cite{dieleman1996direct}, is that junction self-heating shows a strong dependence on the substrate material. Figure~4 compares the IVCs of two four-junction series arrays (each composed of $\phi$2-µm junctions) fabricated on two different substrates: thermally oxidized silicon (1-µm SiO$_2$ surface layer) and crystalline SiO$_2$. In the case of oxidized silicon, the bend appearing at the linear branch of the IVC indicating the phase transition of a parasitic weak link occurs at a lower bias voltage ($K_1$), whereas for crystalline SiO$_2$, the bend appears at a higher voltage ($K_2$). Consistently, the IVC on oxidized silicon shows stronger back-bending at the gap, implying more significant heating.

We also fabricated devices on crystalline silicon substrates (resistivity $> 5$~k$\Omega\cdot$cm) without an oxidized surface, as well as on fused SiO$_2$. The results confirm that crystalline silicon behaves similarly to crystalline SiO$_2$ in exhibiting relatively weak junction heating, while fused SiO$_2$ demonstrates strong heating comparable to oxidized silicon.

These experimental observations cannot be reconciled with the simple analytical theory in~\cite{dieleman1996direct}, which assumes a small Kapitza conductance and predicts that the substrate material should have little effect. Instead, our data indicate that the interfacial thermal conductance is strongly dependent on substrate texture: crystalline substrates provide much higher interfacial conductance and therefore more effective cooling than amorphous substrates.

Simulations support this conclusion. For amorphous substrates, adopting a Kapitza conductance of $5 \times 10^{3}$~W/(K·m$^{2}$), within the range of measured values for Nb/silica interfaces~\cite{abaloszewa2023thermomagnetic}, leads to reasonable heating effects consistent with experiment. However, to reproduce the weaker heating observed on crystalline substrates, the interfacial conductance must be set about three orders of magnitude higher, around $2 \times 10^{6}$~W/(K·m$^{2}$). Although this number should be interpreted only in an order-of-magnitude sense, it clearly demonstrates that crystalline substrates enable far more efficient heat dissipation.

This large difference can be understood in light of above-mentioned Little’s analysis, which emphasizes the role of  mean free paths , and also reflects the fact that the thermal conductivity of crystalline SiO$_2$ is several orders of magnitude higher than that of amorphous SiO$_2$ at cryogenic temperatures. The corresponding simulated temperature profiles for crystalline and amorphous cases are shown in Fig.~\ref{FigSubDep}, with thermal healing lengths of $\eta \sim 1$~µm and $\eta \sim 3$~µm, respectively, the latter consistent with the estimate in~\cite{dieleman1996direct}.

\begin{figure}[tb]
\centering
\includegraphics[width=3.2in,clip]{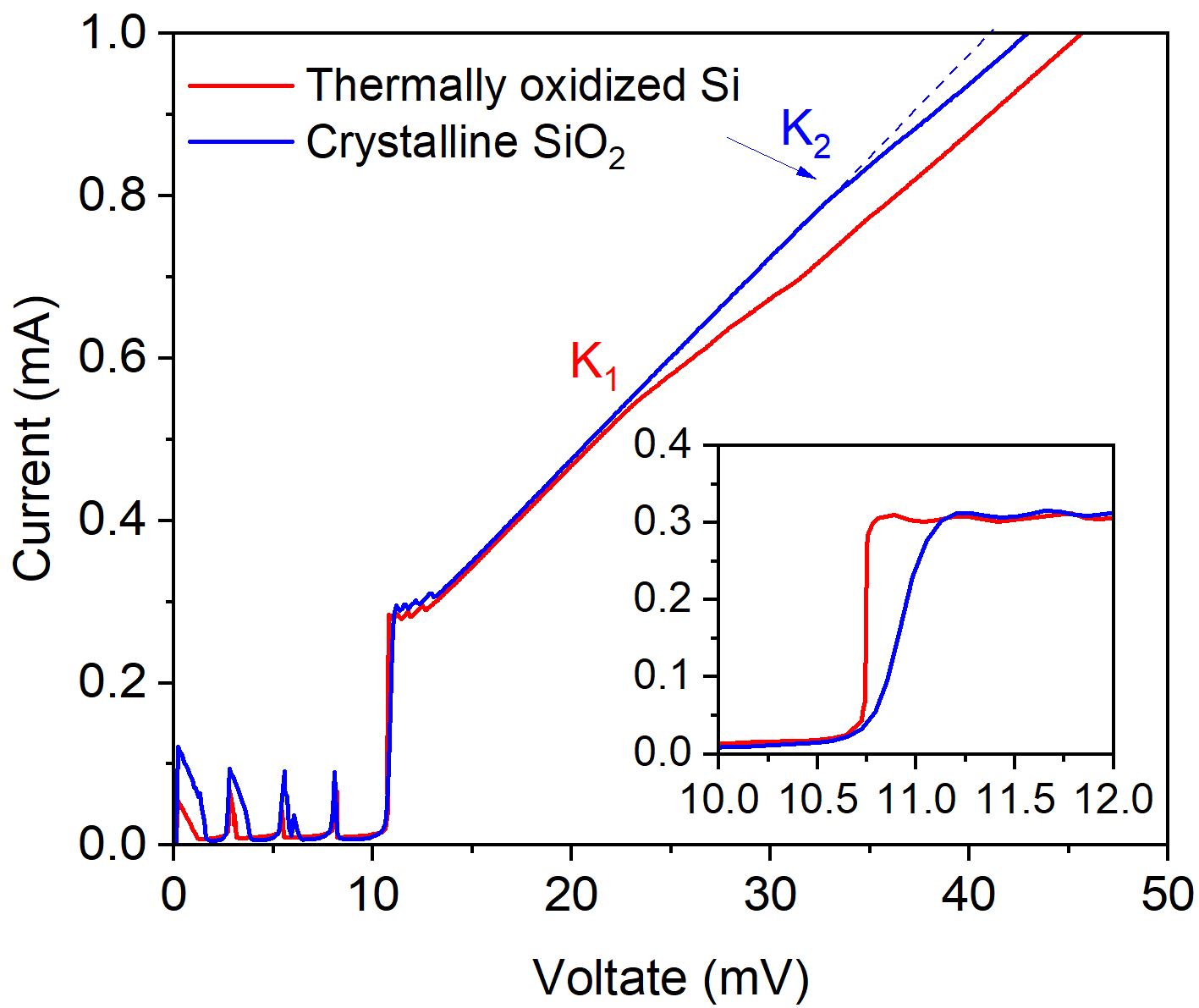}
\caption{IVCs of two four-junction arrays fabricated on oxidized Si (red) and crystalline SiO$_2$ (blue). Bends $K_1$ and $K_2$ are indicated. The blue curve is scaled in current by 0.74 to align with the red one for slope comparison.}
\label{FigSubDepMeasured}
\end{figure}

\begin{figure}[tb]
\centering
\includegraphics[width=3.2in,clip]{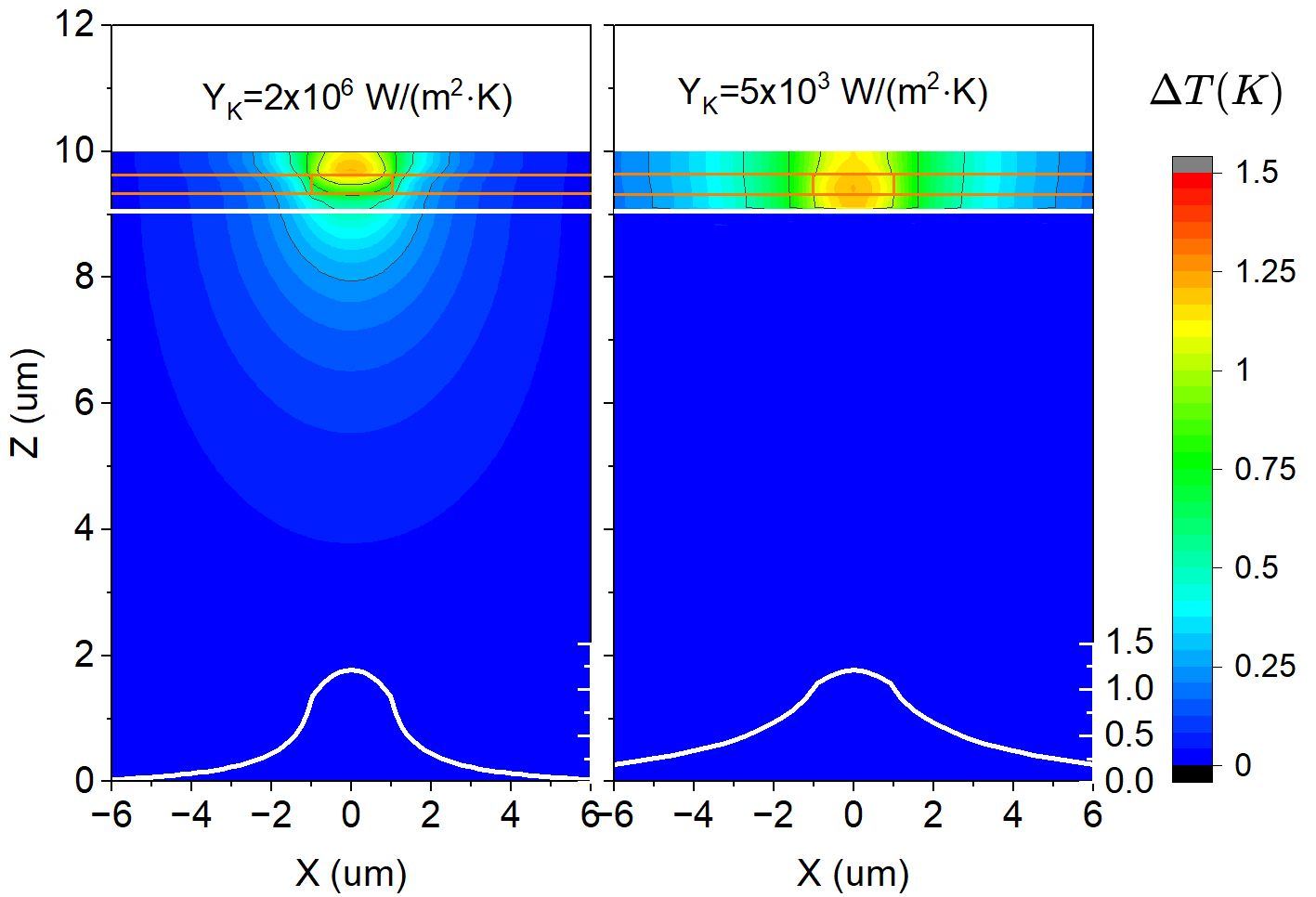}
\caption{Simulated temperature distributions (referring to the bath temperature, $\Delta T=T-T_{bath}$) for a 2-µm junction with (left) large and (right) small Kapitza conductance. The heating power are $\rm{3.6 \,\mu W}$ and $\rm{1 \,\mu W}$ for the two cases to achieve similar temperature at the center of the junctions. Lower panels: temperature decay profiles along a horizontal cross-section.}
\label{FigSubDep}
\end{figure}

\subsubsection{Heating Power Dependence}

\begin{figure}[tb]
    \centering
    \begin{minipage}[t]{0.45\textwidth}
        \centering
        \includegraphics[width=3.2in,clip]{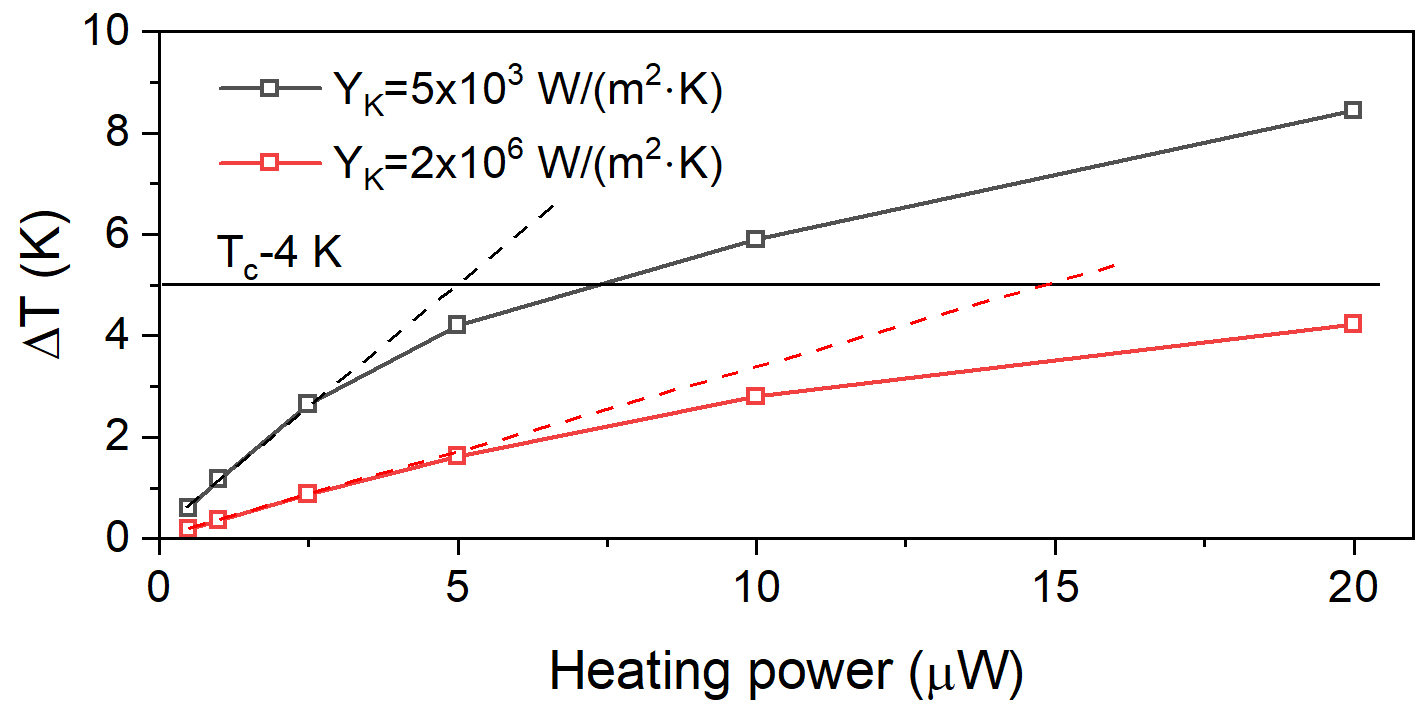}
        \subcaption{(a)}
    \end{minipage}
    \hfill
    \begin{minipage}[t]{0.45\textwidth}
        \centering
        \includegraphics[width=3.2in,clip]{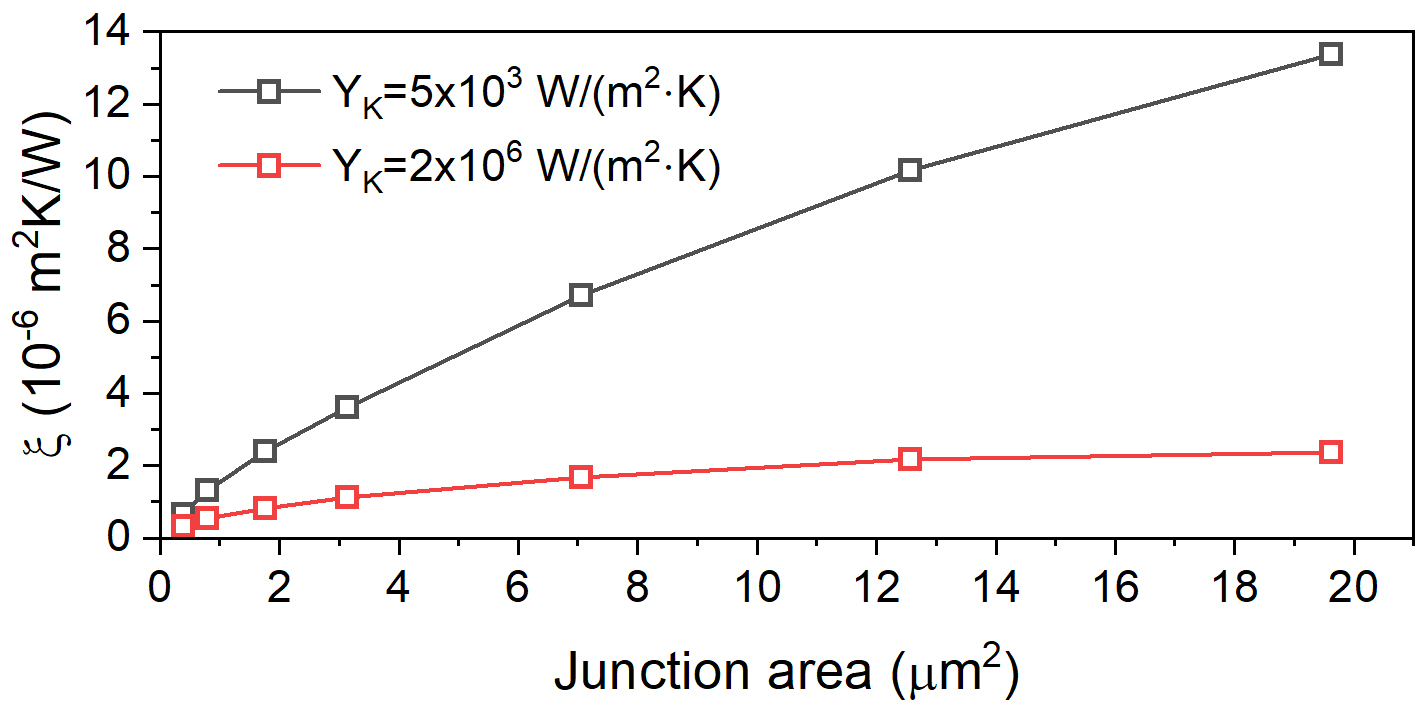}
        \subcaption{(b)}
     \end{minipage}

    \caption{(a) Simulated temperature rise of a 2-µm junction versus heating power for two Kapitza resistances. A line at $\Delta T=T_c-\rm{4\, K}$ indicates the critical temperature when $T_{bath} = \rm{4\, K}$, and the crosses to the heating profiles correspond to the critical heating powers where the junction turns to normal state. (b) Simulated heating coefficient as a function of junction area.}
    \label{FigPowerSizeDep}
\end{figure}

Because the conductivities of the constituent materials increase with temperature, junction self-heating is nonlinear in power. Simulations of a 2-µm junction at different heating powers (Fig.~\ref{FigPowerSizeDep}a) show linear response at low powers, but saturation at higher powers. The dependence can be approximated by
\begin{equation}\label{EqDeltaT}
\Delta T = \eta P e^{-P/P_0},
\end{equation}
where $\eta$ is the small-signal coefficient and $P_0$ a pseudo-saturation power. Junctions reach $T_c$ at a critical power where $\Delta T + T_{\text{bath}} = T_c$. Ignoring this nonlinearity leads to overestimation of junction heating.

\subsubsection{Junction Size Dependence}

\begin{figure}[tb]
    \centering
    \begin{minipage}[t]{0.45\textwidth}
        \centering
        \includegraphics[width=3.2in,clip]{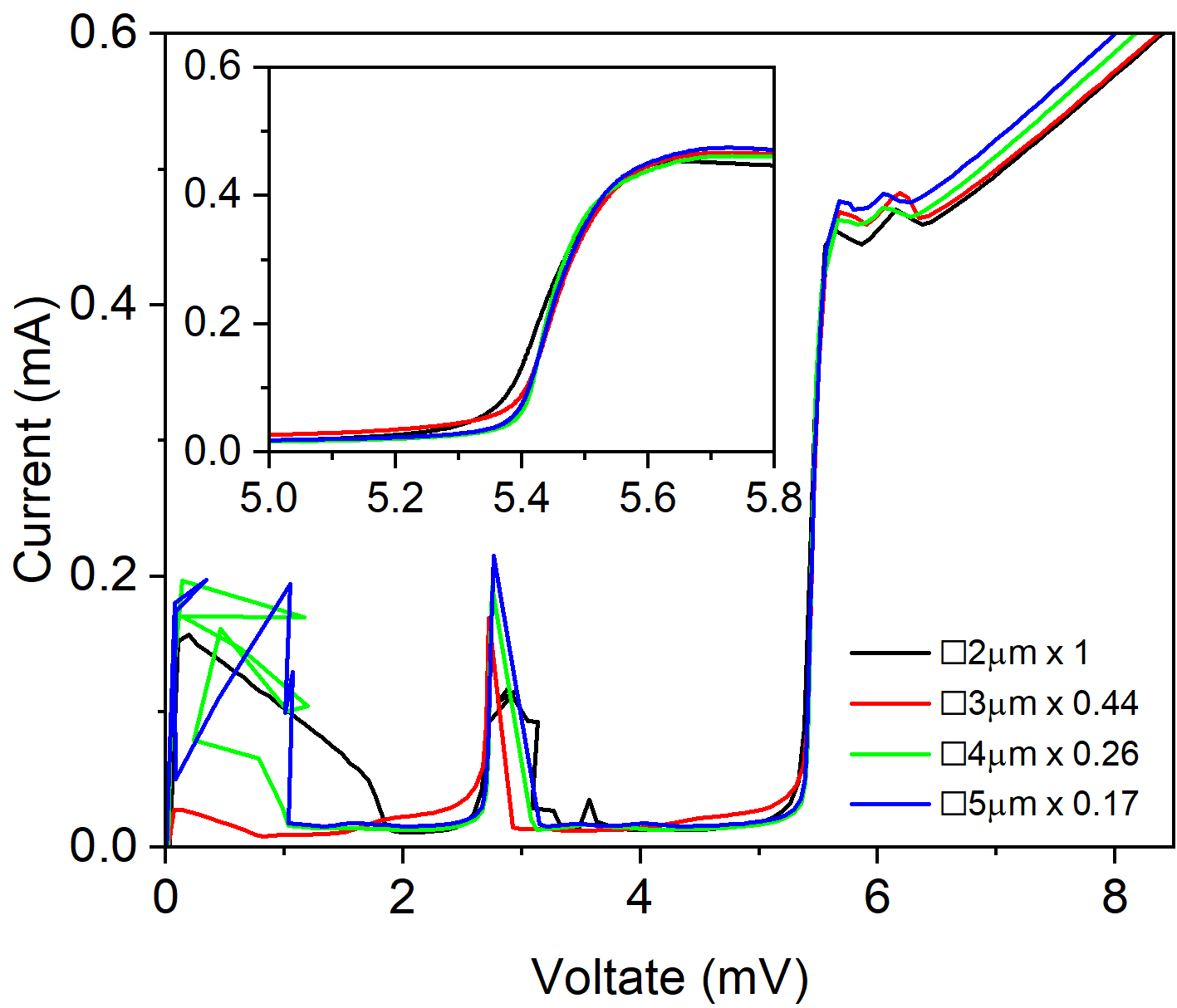}
        \subcaption{(a)}
    \end{minipage}
    \hfill
    \begin{minipage}[t]{0.45\textwidth}
        \centering
        \includegraphics[width=3.2in,clip]{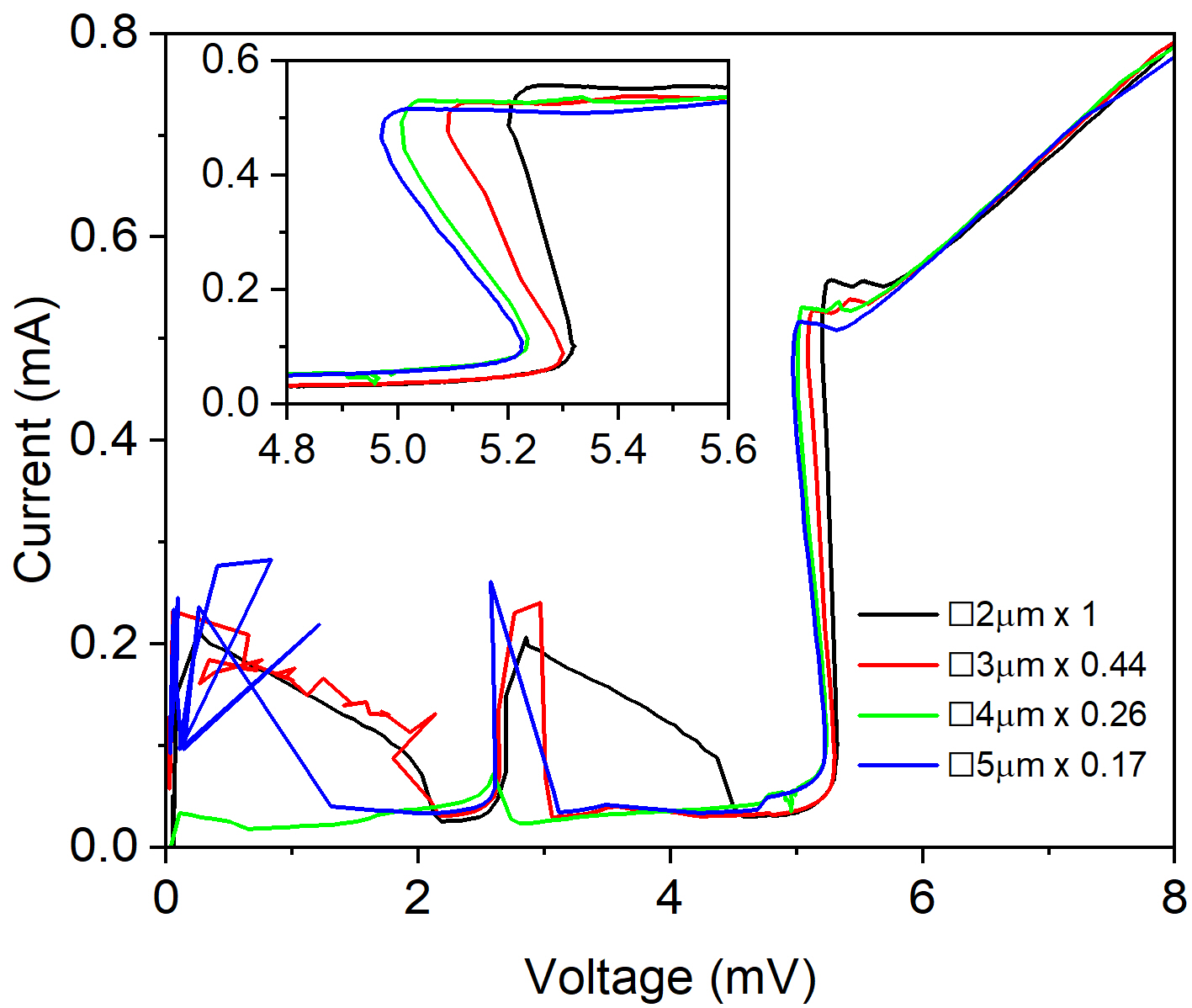}
        \subcaption{(b)}
    \end{minipage}

    \caption{IVCs of four two-junction series arrays with junction lateral sizes of 2, 3, 4, and 5 µm fabricated on (a) crystalline  SiO$_2$ substrate and (b) thermally oxidized Si substrate. Insets show the bending features near the gap. The IVCs are normalized by multiplying the current with a factor inversely proportional to the junction area so that they can be compared at the same current as the 2-µm junction. }
    \label{FigSizeMeas}
\end{figure}

Experiments show strong junction size dependence for devices on amorphous substrates, with larger junctions exhibiting stronger back-bending. This dependence is weak on crystalline substrates (Fig.~\ref{FigSizeMeas}). Simulations reproduce this effect.

We define a heating coefficient
\begin{equation}\label{EqHeatingCoefficient}
\xi = \frac{\Delta T}{P_h/A_J},
\end{equation}
where $A_J$ is junction area and $P_h$ the heating power. If junction cooling is dominated by vertical (downward) heat diffusion while lateral diffusion is negligible, the heating coefficient will be independent of junction area. However, when lateral diffusion becomes relevant, the effective cooling region extends to an area with approximate diameter $\phi_J + 2\eta$, where $\eta$ is the thermal healing length and $\phi_J$ the junction diameter.

Two limiting cases can be distinguished. If $\eta$ is small compared to the junction size, both heating power and cooling capacity scale linearly with junction area, and thus the heating coefficient shows only weak size dependence. In contrast, if $\eta$ exceeds the junction size, the cooling power depends only weakly on area, while the heating power remains proportional to junction area. In this regime, the heating coefficient increases approximately linearly with junction area.

This behavior is illustrated in Fig.~\ref{FigPowerSizeDep}(b) and is consistent with the experimental trends observed in Fig.~\ref{FigSizeMeas}.

\subsubsection{Series Arrays on Islands}
\begin{figure}[tb]
    \centering
    \begin{minipage}[t]{0.45\textwidth}
        \centering
        \includegraphics[width=3.2in,clip]{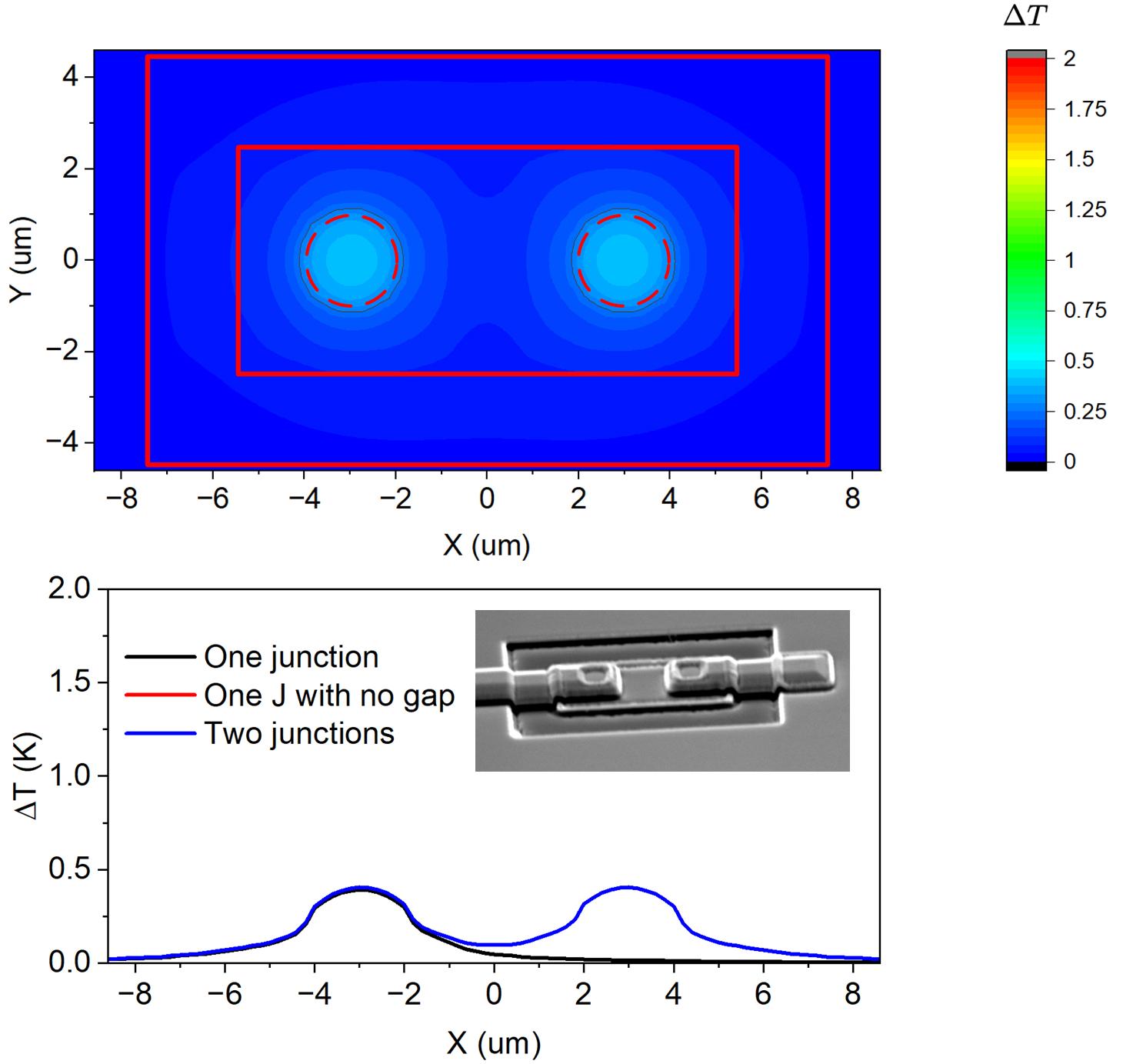}
        \subcaption{(a)}
    \end{minipage}
    \hfill
    \begin{minipage}[t]{0.45\textwidth}
        \centering
        \includegraphics[width=3.2in,clip]{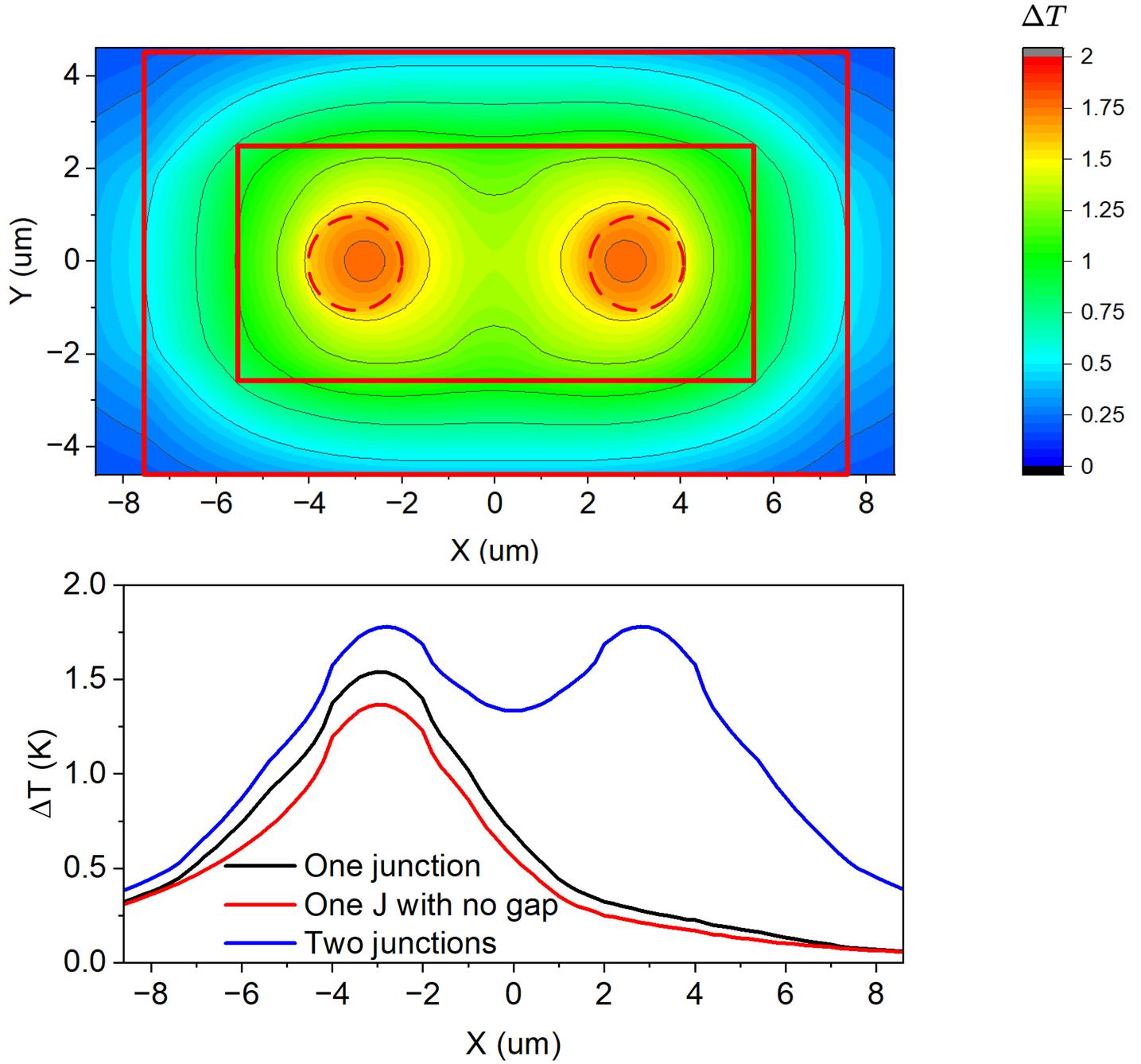}
        \subcaption{(b)}
    \end{minipage}

    \caption{Self-heating of a $\phi_J$ = 2 µm junction pair on an island separated from the outer Nb ground plane at two cases: (a) on a crystalline SiO$_2$ substrate and (b) on a thermally oxidized Si substrate. A power of 1 µW is applied to each junction when heated. The inset in (a) is an SEM image of a junction pair. }
    \label{FigIsland}
\end{figure}

Series junction arrays are often adopted in SIS mixer designs to extend the dynamic range of response. Such arrays are typically formed by connecting junction pairs, with each pair located on a small Nb island that is electrically isolated from the ground plane, as shown in the inset of Fig.~\ref{FigIsland}(a). The interruption of the ground plane impedes lateral heat diffusion, particularly when the island dimensions are comparable to or smaller than the thermal healing length. As a result, junctions placed on islands may experience stronger self-heating than those connected to a continuous ground plane.

To investigate this effect, we performed simulations for junction pairs patterned on crystalline and amorphous substrates, using island geometries identical to those in our test devices, such as those shown in Fig. \ref{FigSubDepMeasured} and Fig.\ref{FigSizeMeas}. Each island measured $5 \times 10$~µm$^2$ with a surrounding gap of 1~µm. In the high Kapitza conductance case (crystalline substrate), where the healing length is about 1~µm, the island size is much larger than $\eta$, and thus the island has little influence on the junction cooling, as seen in Fig.~8(a). In contrast, in the low Kapitza conductance case (amorphous substrate), $\eta$ becomes comparable to the island width, and the confinement effect leads to a noticeable increase in junction temperature, on the order of 10\%. Furthermore, when two junctions on the same island (spaced by 6~µm) are biased simultaneously, the local heating effects combine, causing an additional $\sim$10\% rise in temperature. Experimentally, we observed that junction arrays exhibited slightly stronger self-heating than isolated test junctions fabricated on larger continuous ground planes on the same wafer. This qualitative agreement with the simulations supports the interpretation that restricted lateral diffusion on small islands, combined with thermal interaction between neighboring junctions, enhances junction heating in series arrays.

\subsubsection{Cooling With Al Cap Layer}

Due to the high thermal conductivity of aluminum at cryogenic temperatures, aluminum layers can provide remarkable enhancement in junction cooling. This effect was already noted in~\cite{dieleman1996direct}, where aluminum wiring layers were shown to improve heat spreading. In our case, the SIS mixer chips under study are relatively large (12~mm $\times$ 10~mm)~\cite{ezaki2019fabrication}, which provides sufficient area to accommodate additional microfabrication processes. Taking advantage of this, an aluminum cap layer with a thickness of 300~nm was deposited on top of the Nb wiring layer and patterned using a laser writer followed by a lift-off process. A clear reduction of junction self-heating was observed in the measured IVCs, as shown in Fig.~\ref{FigAlPatchMeasurement}.

\begin{figure}[tb]
\centering
\includegraphics[width=3.2in,clip]{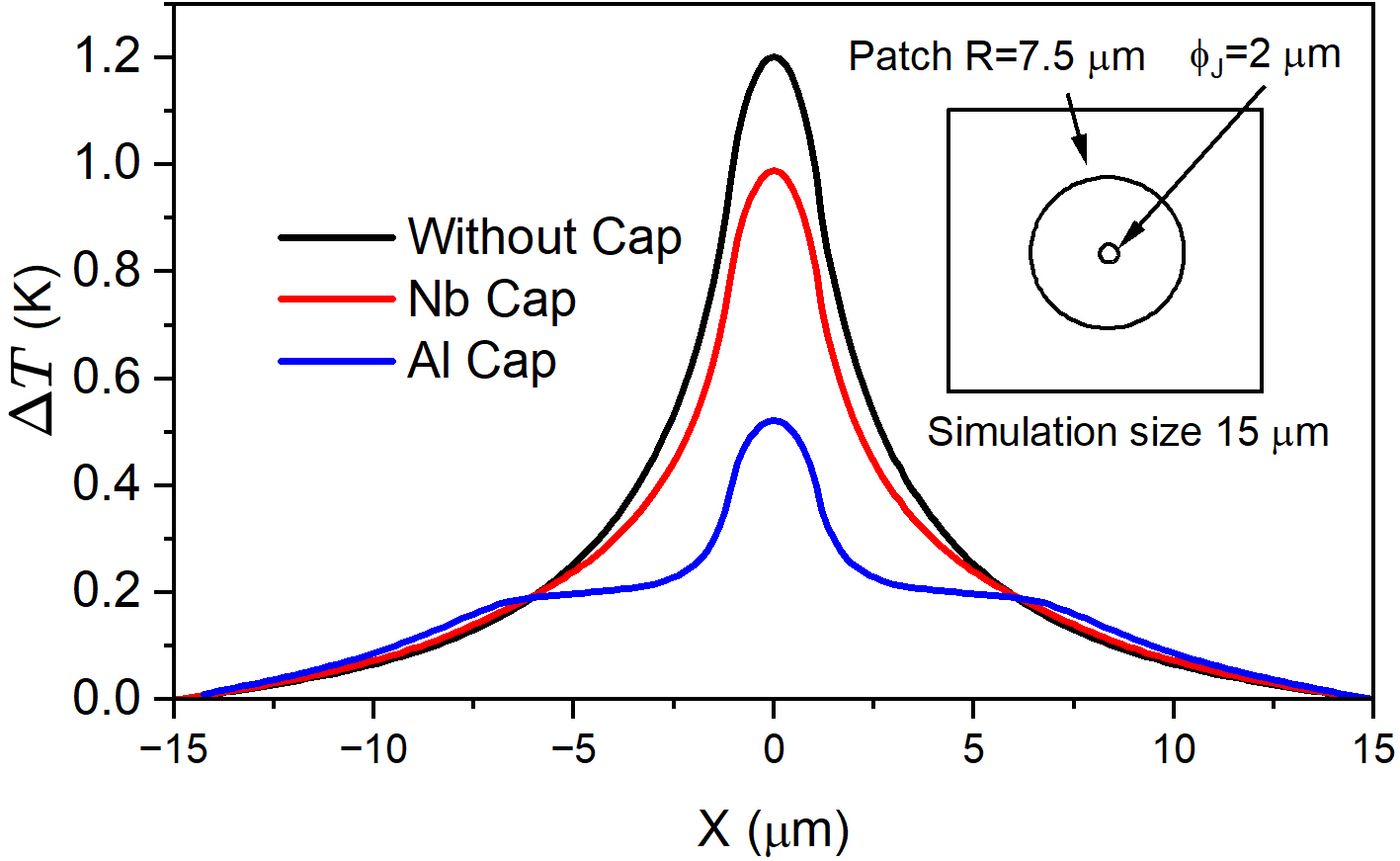}
\caption{Cooling enhancement with an Al cap layer and comparison with an Nb cap layer and with no cap layer. Temperature distributions along a horizontal line through the junction center are shown for the three cases. The junction and patch sizes are 2 µm and 15 µm in diameter, respectively, and the thickness of the cap layers is 300 nm. The junction is fabricated on a thermally oxidized substrate and the heating power is set to 1 µW.}
\label{FigCap}
\end{figure}

This improvement is reproduced by simulation, as illustrated in Fig.~\ref{FigCap}. Three conditions were compared: no cap layer, a Nb cap layer, and an Al cap layer, all with identical geometry (15~µm diameter and 300~nm thickness). The results show that the Al patch efficiently spreads heat laterally and transfers it downward into the substrate, reducing the junction temperature by approximately 60\% relative to the no-cap case. By contrast, a Nb patch of the same size and thickness lowers the temperature by only about 20\%.

\subsubsection{Wiring pattern}
According to above simulation results the wiring Nb only weakly affect the junction cooling. This is because this layer is separated from the base Nb with 300 nm thick sputtered $\rm{SiO_2}$, which has a thermal conductivity of an order of magnitude lower than Nb at 4 K. A weak cooling enhancement of Nb cap layer was also observed in experiments. For the same reason, the pattern shape of Nb wiring only show very minor influence on junction heating in our simulation.

\section{Junction Heating Exhibited in Current–Voltage Characteristics}

\subsection{Modeling Method}

The temperature distribution obtained from finite-element analysis cannot be compared directly with experimental data. To enable quantitative comparison, we introduce a heating coefficient, defined in Eq.~(\ref{EqHeatingCoefficient}), that relates the power dissipation to the junction temperature rise. This coefficient is then linked to the bending angle of the IVCs at the gap voltage. By combining this heating coefficient with the known temperature dependence of the gap voltage and gap broadening, IVCs that incorporate self-heating can be calculated.

IVCs are calculated by using the semiconductor tunneling model:
\begin{equation}\label{EqIV}
I \propto \int_{-\infty}^{\infty} N_L(E) N_R(E+eV) \left[f_L(E) - f_R(E+eV)\right] dE,
\end{equation}
where $N(E)$ is the quasiparticle density of states (DOS) and $f(E)$ the Fermi distribution. Subscripts $L$ and $R$ denote the left and right electrodes, respectively, and $E$ is the electron energy. According to BCS theory, the DOS is
\begin{equation}\label{EqDOS}
N(E) = \frac{E}{\sqrt{E^2 - \Delta(T)^2}},
\end{equation}
where the superconducting energy gap $\Delta$ depends on temperature. Its approximate temperature dependence is expressed as
\begin{equation}\label{EqGapT}
\frac{\Delta(T)}{\Delta(0)} = 1 - \left(\frac{T}{T_c}\right)^3 \left(1 - \sqrt{\frac{T}{T_c}}\right).
\end{equation}
Measured gap voltages agree well with this model.

Gap broadening, associated with finite quasiparticle lifetime, is also temperature dependent. It can be introduced either by adding an imaginary part to $E$~\cite{dynes1978direct} or to $\Delta$~\cite{mitrovic2007correct} in Eq.~(\ref{EqDOS}). We adopt the former approach, since it yields a smoother transition of the DOS across $T_c$:
\begin{equation}\label{EqDOSCoxplex}
N(E) = \text{Re} \left[ \frac{E+i\epsilon}{\sqrt{(E+i\epsilon)^2 - \Delta^2}} \right],
\end{equation}
where $\epsilon$ is fitted from measured broadening as a function of temperature.

Because both $\Delta$ and $\epsilon$ now depend on junction temperature, the model must be solved self-consistently. The effective junction temperature is given by
\begin{equation}\label{EqTCal}
T = T_{\text{bath}} + \xi \frac{VI}{A_J},
\end{equation}
where $\xi$ is the heating coefficient, $A_J$ the junction area, and $V$ and $I$ the applied bias voltage and current. Equation (\ref{EqIV}) thus becomes an implicit function that is solved iteratively. The procedure begins with an assumed junction temperature (typically $T_{\text{bath}}$), from which current is calculated using Eqs.~(\ref{EqIV})-(\ref{EqDOSCoxplex}). The updated temperature is then obtained via Eq.~(\ref{EqTCal}), and the process repeats until convergence.

For IVCs with back-bending, Eq.~(\ref{EqIV}) becomes multivalued with respect to $V$, and the negative-resistance branch is not accessible in the forward calculation. In this case, the inverse problem must be solved by iterating voltage from a given current.

\subsection{Bending Features in IVCs and Comparison With Experiment}

To characterize junction heating, we define a critical bias voltage $V_c$ at which the electrodes are heated to $T_c$. This definition is convenient, since a clear change in resistance is typically observed around this voltage when a weak link is present near the junction. The heating coefficient can then be expressed as
\begin{equation}\label{EqHeatingCoefficientXi}
\xi = \frac{\Delta T A_J}{P_h} = \frac{(T_c - T_{\text{bath}}) A_J}{(\bar{V}_c V_{\text{gap}})^2 / R_n}
\approx \frac{3.75}{\bar{V}_c^2 V_{\text{gap}} J_c},
\end{equation}
where $\bar{V}_c = V_c / V_{\text{gap}}$, $R_n$ is the normal resistance, and $J_c$ is the critical current density.

For junctions with $J_c = 10$~kA/cm$^2$ (typical in our experiments), $\bar{V}_c = 2$ and 3 correspond to heating coefficients of $3.5 \times 10^{-6}$ and $1.5 \times 10^{-6}$~m$^2$K/W, respectively, consistent with simulation results in Fig.~\ref{FigSizeMeas}(b). These values also correspond to the bias points at which unexpected bends are observed in measured IVCs for junctions on amorphous and crystalline substrates.

\begin{figure}[tb]
\centering
\includegraphics[width=3.2in,clip]{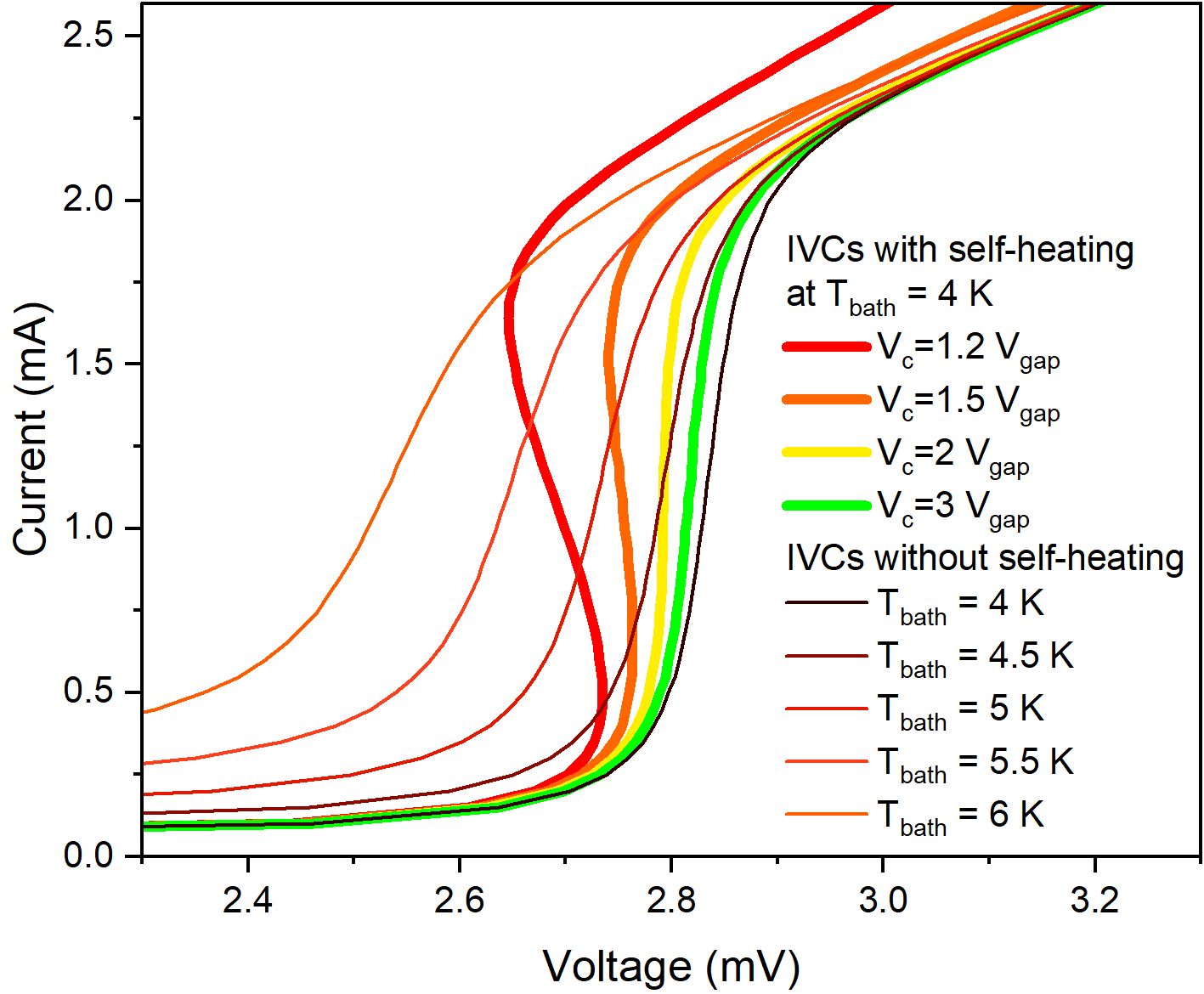}
\caption{Bending feature of IVCs (thick lines) at the gap voltage simulated at various heating coefficients and at a bath temperature of 4 K. A series of IVCs (thin lines) free of self-heating and calculated at various bath temperatures are plotted in the background to trace the actual temperature of the foreground IVCs.}
\label{FigBackbendingSim}
\end{figure}

Simulated IVCs at various heating coefficients are shown in Fig.~\ref{FigBackbendingSim}. Bending angle near the gap voltage rotates backward with increasing the heating coefficient. For comparison, IVCs of junctions without self-heating are also plotted at different bath temperatures. These provide a “coordinate system” from which the effective junction temperature of self-heated IVCs can be determined by overlaying data.

\begin{figure}[tb]
\centering
\includegraphics[width=3.2in,clip]{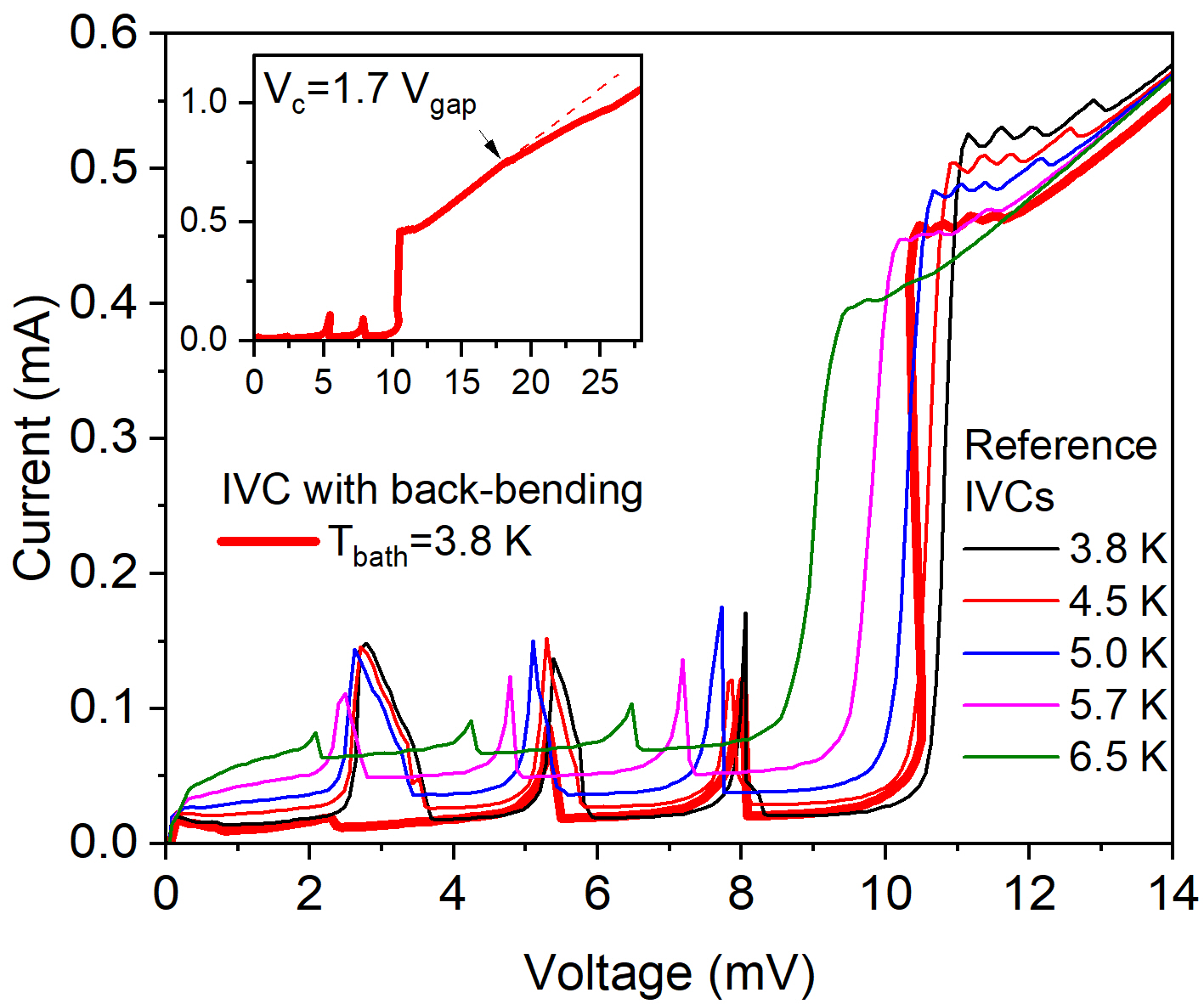}
\caption{Measured IVC with back-bending due to self-heating. It is overlapped with a series of IVCs of a less self-heated junction measured at various bath temperatures tuned by using a heater anchored on the cold stage. The inset shows the same IVC over a wider voltage range, from which the critical voltage can be estimated to be 1.7 $V_{\text{gap}}$. The reference IVCs are enlarged by a factor of 1.3 in current to align with the IVC showing back-bending.  }
\label{FigBackbendingMeas}
\end{figure}

This method can also be applied experimentally. Figure~\ref{FigBackbendingMeas} shows the IVC of a four-junction series array fabricated on thermally oxidized silicon. Strong back-bending near the gap is evident. On the same plot, we overlay IVCs from another device fabricated on crystalline SiO$_2$, measured at bath temperatures varied by a heater. The cross points allow estimation of junction temperature in the self-heated device. At the upper turning point, just above the gap, the temperature rise is about 1.5~K (to 5.5~K). This corresponds to $\bar{V}_c \approx 1.5$, close to the $\bar{V}_c \approx 1.7$ estimated directly from the bending voltage (inset of Fig.~\ref{FigBackbendingMeas}).

Note that the measured IVCs also show excess current (knee structure) due to proximity effect, not included in the model, which may introduce uncertainty. Nonetheless, the consistency of these rough estimates is sufficient to explain the observed phenomena.

\subsection{Critical Heating Voltage and Comparison With Measurement}

\begin{figure}[tb]
\centering
\includegraphics[width=3.2in,clip]{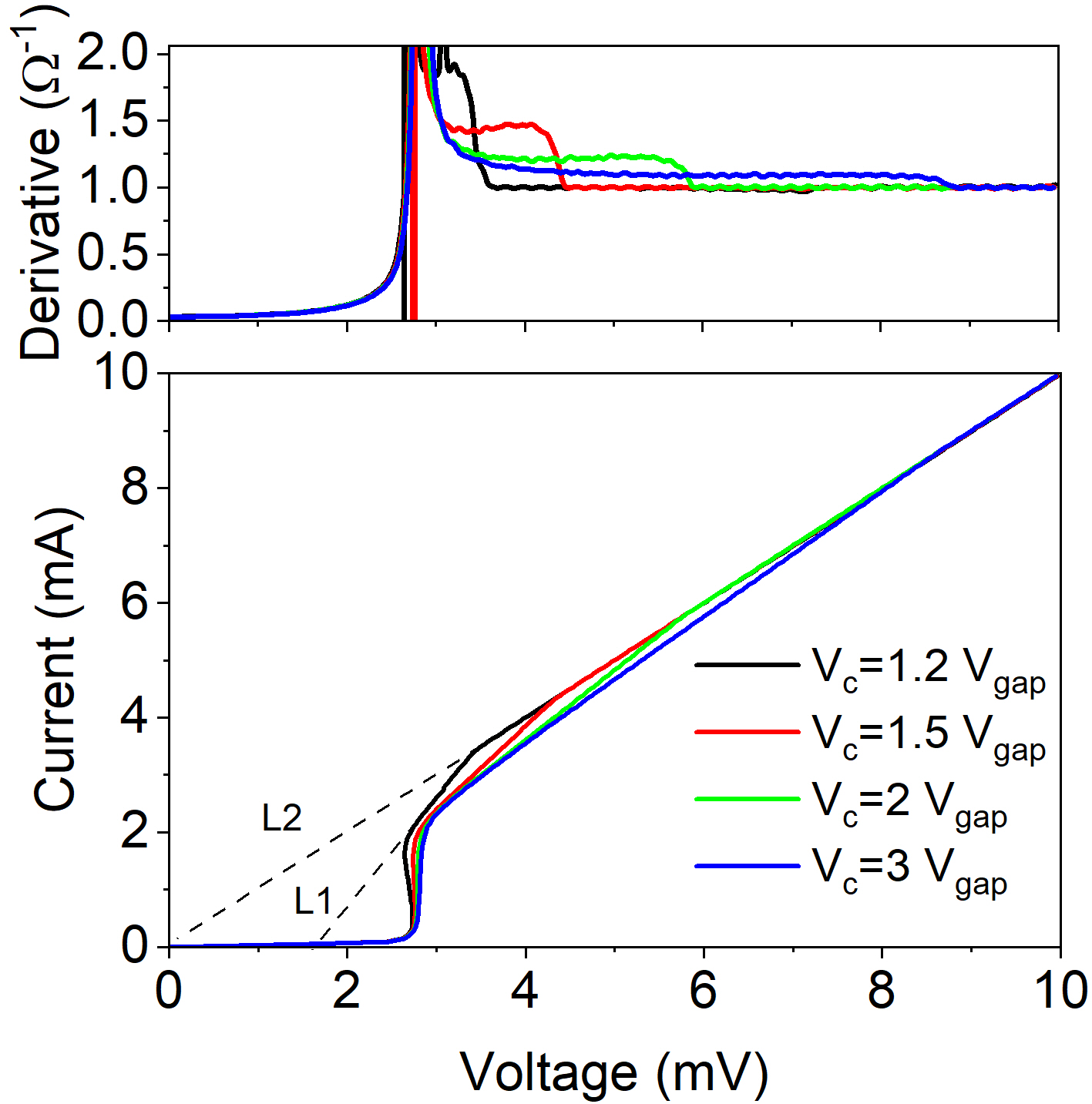}
\caption{Simulated IVCs (lower panel) at $T_{\text{bath}}$ = 4 K for a self-heated junction corresponding to four typical critical voltages. In addition to the manifestation of self-heating at the gap voltage, discontinuities of resistance at the linear branch at $V_c$ are observed. The upper panel shows the derivatives of the IVCs.  }
\label{FigDistantBendSimu}
\end{figure}

Even without incorporating weak links, simulated IVCs show features at the critical voltage $V_c$. As shown in Fig.~\ref{FigDistantBendSimu}, discontinuities appear in the derivative of the IVCs at $V_c$, especially for large heating coefficients (small $V_c$). These arise because the DOS function is temperature dependent below $T_c$ but temperature independent above $T_c$.

In simulations, the extrapolated linear branch beyond $V_c$ passes through the origin, while the pre-$V_c$ portion intersects the voltage axis at a finite value. This behavior is particularly visible when $\bar{V}_c \approx 1.2$.

\begin{figure}[tb]
\centering
\includegraphics[width=3.2in,clip]{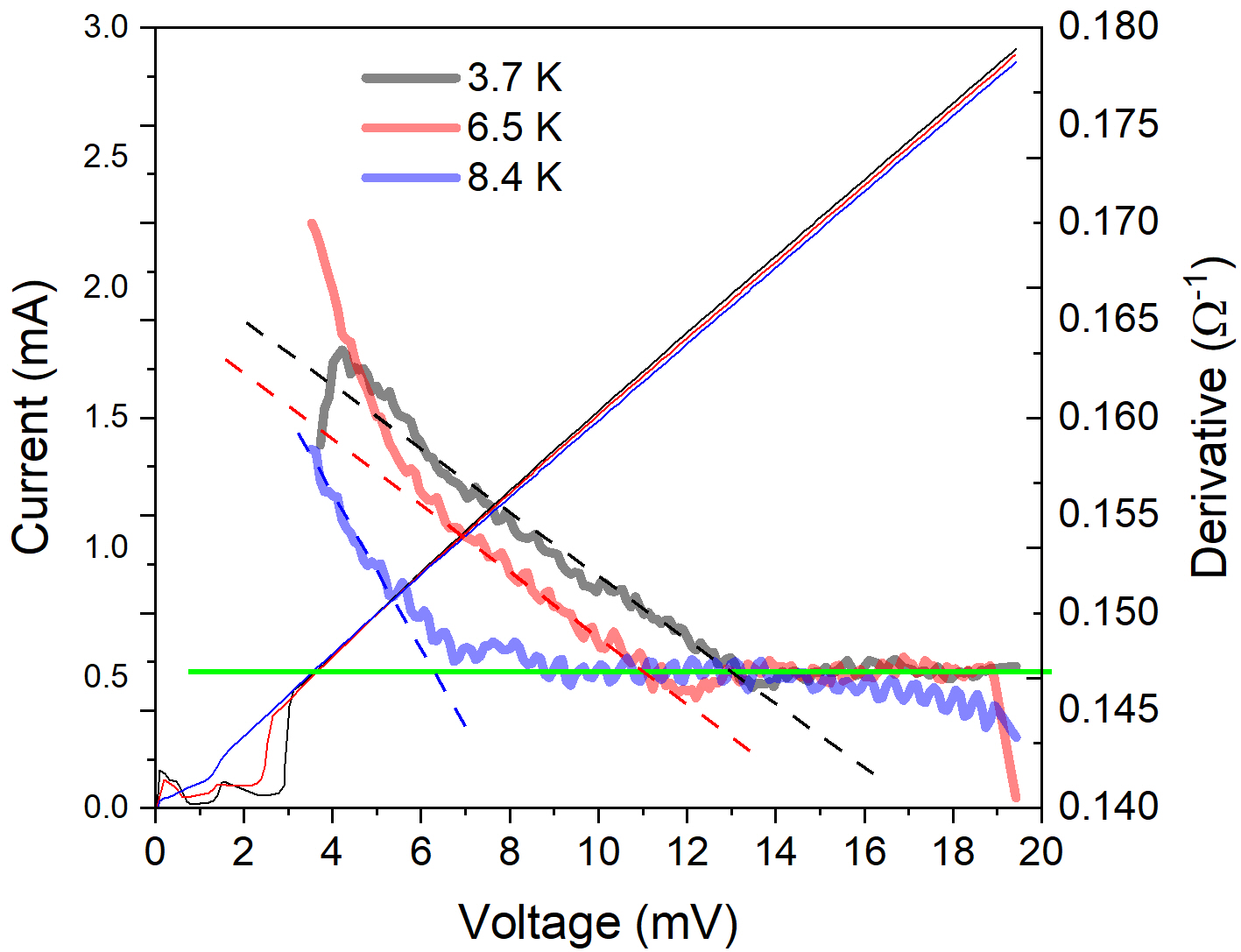}
\caption{Measured IVCs and their derivatives of a 2-µm square SIS junction at three bath temperatures. The junction was fabricated with the self-aligned lift-off method on an SOI wafer, and the current density is about 9 kA/cm$^2$.  }
\label{FigDistantBendMeas}
\end{figure}

Experimentally, some junctions fabricated using a self-aligned lift-off process do not show obvious bends in their IVCs (Fig.~\ref{FigDistantBendMeas}). However, their derivatives reveal two distinct regions: a lower-voltage region with temperature-dependent resistance and a higher-voltage region with constant resistance. The transition voltage between these regions decreases with increasing bath temperature, consistent with heating effects.

In these devices, the transition occurs at $\sim 4V_{\text{gap}}$, higher than the $V_c\sim 3V_{\text{gap}}$ estimated with junctions showing weak links. This can be explained by considering that the bend occurs not when the junction reaches $T_c$, but when the weak link critical current $I_c(T_b)$ is reduced to the bias current $I_b$. With $I_b = I_c(0) [1-(T_b/T_c)^n]$ and $n=1$ as determined experimentally (see Sec.~IV), the transition temperature $T_b$ is generally below $T_c$. Only when $I_c(0) \gg I_b$ does $T_b$ approach $T_c$.

\section{Weak Links}
\subsection{Possible Location of Weak Links}

\begin{figure}[tb]
\centering
\includegraphics[width=3.2in,clip]{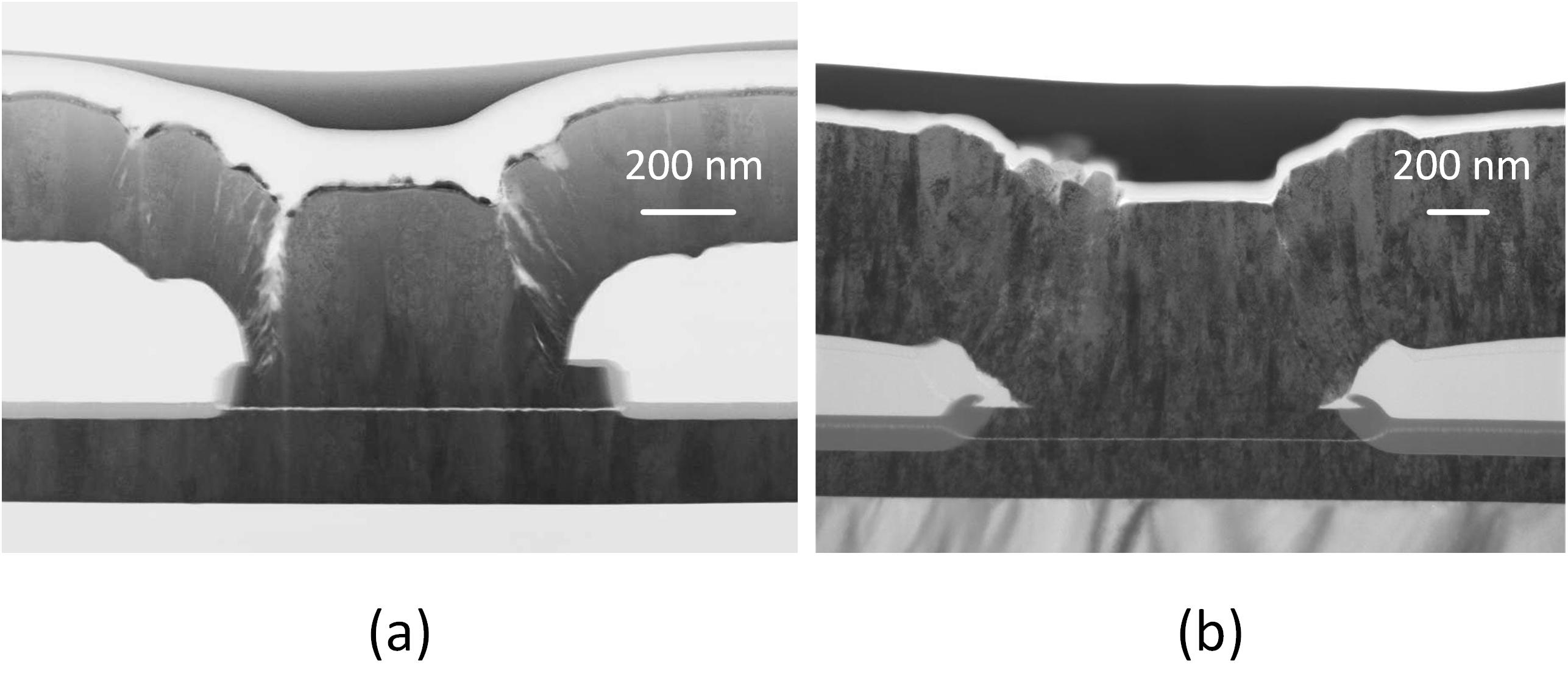}
\caption{TEM images of two junctions fabricated by (a) the machine-aligned etching process and (b) the self-aligned lift-off process. The left image exhibits a ring of cracks around the contacting hole, while the right one does not. }
\label{FigTEMCrack}
\end{figure}

For weak links to influence IVCs at $\bar{V}_c \sim 2$–3, they must be located within the thermal healing length of the SIS junction (about 1~µm on crystalline substrates). Transmission electron microscope (TEM) images provide evidence. Figure~\ref{FigTEMCrack} compares two junctions: one fabricated using the machine-aligned etching process and another by self-aligned lift-off. The etched junction shows a ring of crack around the contact hole, while the lift-off junction shows a continuous wiring layer.

From the texture directions in the TEM images, Nb grains tend to grow perpendicular to the underlayer. Growth is likely interrupted at sharp turning edges. Since contact windows formed by etching are sharper than those formed by lift-off, cracks are more likely in the etched process, leading to weak links at via holes.

\subsection{Direct Measurement of Weak Links}

\begin{figure}[tb]
\centering
\includegraphics[width=3.2in,clip]{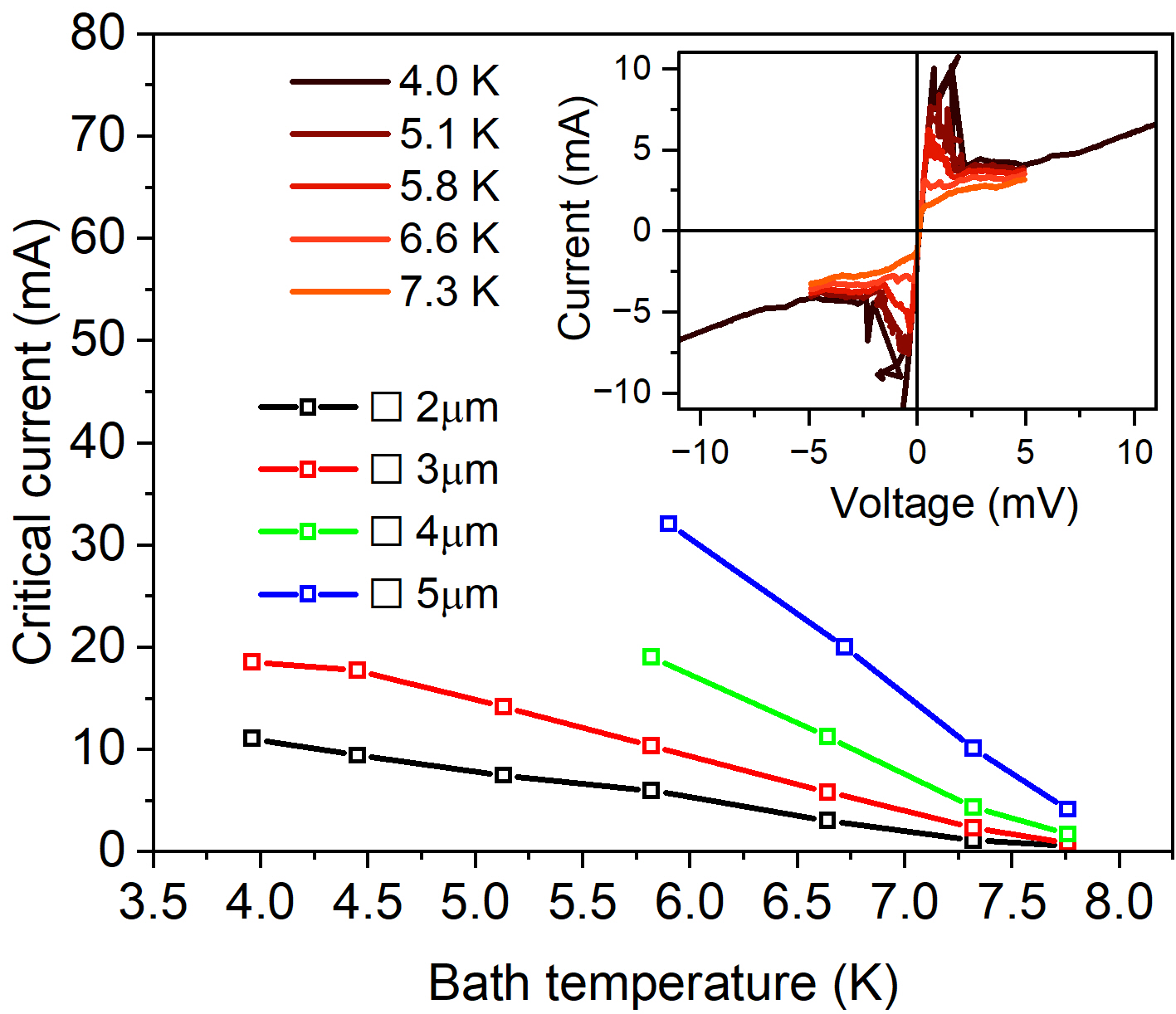}
\caption{Critical currents of weak links parasitically cohabiting with dummy SIS junctions of various sizes as a function of bath temperature. The inset shows IVCs of a 2-µm junction measured at various bath temperatures. A small contact resistance is left uncorrected.}
\label{FigIcTOxFree}
\end{figure}

To isolate weak-link behavior, we fabricated devices without junction oxidation, so that the SIS junctions were effectively shorted. In this case, the IVC reflects only the weak link. Figure~\ref{FigIcTOxFree} shows the IVCs of a 2-µm dummy junction measured at various bath temperatures. A zero-resistance current is sustained up to a critical current, beyond which a finite resistance of about 2~$\Omega$ appears, very close to the resistance increase observed at $K_1$ in Fig.~\ref{FigAlPatchMeasurement}.

At 4~K, the weak-link critical current is about 10~mA, beyond the range typically used in SIS mixer operation. However, the critical current decreases nearly linearly with temperature and approaches values comparable to those of SIS junctions when heated near $T_c$. Thus, weak links can routinely manifest in IVCs if they form near SIS junctions.

It should be noted that weak links may also form elsewhere in mixer circuits, not necessarily adjacent to SIS junctions. Such weak links, however, usually remain undetected because they are not strongly heated and thus maintain large critical currents outside the measurement range.

\subsection{Sensitivity to Annealing}
Weak links are found to be sensitive to thermal treatment. Annealing at 200$^\circ$C for 30 minutes reduces the critical current of a 2-µm dummy junction from about 10~mA to less than 1~mA at 4~K. Similar behavior was reported in~\cite{imamura2002bias}. This suggests that weak links are not stable structural features but are associated with grain boundaries or cracks that can undergo irreversible changes upon moderate thermal cycling.

Weak links can be suppressed by process improvements. For example, bias sputtering has been shown to strengthen Nb films~\cite{imamura2002bias}, while chemical-mechanical polishing to flatten the surface prior to final wiring deposition is also likely to mitigate weak-link formation.

\section{Impact on SIS Mixer Performance}

\subsection{Impact of Temperature Rise}

Self-heating can degrade SIS mixer performance through two mechanisms: (i) the generation of excess thermally stimulated quasiparticles and (ii) the reduction of nonlinearity, which lowers conversion gain at elevated temperatures. The magnitude of temperature rise is proportional to the total heating power, which is the sum of dc power and LO power.

As an example, consider a 2-µm junction with $J_c = 10$~kA/cm$^{2}$. At 140~GHz and 500~GHz, the heating powers are about 0.4~µW and 1.5~µW, respectively. For junctions fabricated on crystalline SiO$_2$ or silicon substrates, the heating coefficient is approximately $0.4 \times 10^{-6}$~m$^2$K/W from simulation. This yields temperature rises of about 0.2~K and 0.6~K, respectively. While 0.2~K is negligible at a bath temperature of 4~K, a 0.6~K rise is marginal. Since a full analysis of performance degradation is beyond the scope of this work, no further discussion is given here.

\subsection{Impact of Weak Links}

Weak links, if present at contact holes, primarily affect SIS mixer performance through their parasitic inductance~\cite{shan2004anomalous}. The Josephson inductance of a weak link is
\begin{equation}
L_J \approx \frac{\hbar}{2e I_J} \approx \frac{\hbar}{2e \eta I_c},
\end{equation}
where $I_J$ is the weak-link critical current, $I_c$ the SIS junction critical current, $e$ the electron charge, $\hbar$ the reduced Planck constant, and $\eta = I_J/I_c$. The corresponding reactance at frequency $f$ is
\begin{equation}
\omega L_J = \frac{f}{f_{\text{gap}}} \frac{R_n}{1.5 \eta},
\end{equation}
where $R_n$ is the junction normal resistance and $f_{\text{gap}}$ the gap frequency.

In a simple LC tuning circuit, the effect of this parasitic inductance can be estimated by comparing it with the designed tuning inductance $L_T$:
\begin{equation}
\frac{L_J}{L_T} = \frac{f}{f_{\text{gap}}} \frac{Q}{\eta},
\end{equation}
where $Q = \omega_0 R_n C \approx 0.75 V_{\text{gap}} C_J \omega_0/I_c$, $C_J$ is junction capacitance, and $\omega_0 = 1/\sqrt{L_T C_J}$ the tuning resonance frequency.

The additional inductance shifts the resonance frequency by $\Delta f \approx L_J/(2L_T)$ if $L_J \ll L_T$. For a practical example with $f_0 = 140$~GHz, $\eta \approx 20$, $J_c = 10$~kA/cm$^2$, $C_f = C_J/A_J = 60$~fF/µm$^2$, and $V_{\text{gap}} = 650$~GHz, we obtain $Q \approx 1$ and $L_J/L_T \approx 0.01$, which is negligible.

In the worst case, if annealing reduces the weak-link critical current to one-tenth of its original value, $L_J/L_T$ increases tenfold to $\sim 0.1$, corresponding to a 5\% reduction in resonance frequency. Although still moderate, this effect could become significant at higher operating frequencies or lower current densities.

\section{Conclusion}

The self-heating of SIS tunnel junctions has been studied using finite-element analysis and IVC modeling, and the results were compared with experiments. The simulations account for the strong temperature dependence of thermal conductivities, leading to nonlinear heating versus power characteristics.

Experimental evidence shows that interfacial thermal resistance between the Nb base electrode and the substrate strongly depends on substrate texture. Crystalline substrates provide significantly lower resistance than amorphous ones, consistent with Little’s theory, and result in weaker heating. Junction heating also depends on junction size and wiring pattern. Adding an aluminum cap layer on top of the Nb wiring was found to enhance cooling significantly, consistent with simulation.

The simulated heating coefficient was incorporated into IVC models, enabling reproduction of bending features at the gap voltage comparable to experiment. Another key measurable feature is the critical voltage $V_c$, where junctions are heated to $T_c$. In devices with weak links near the junction, $V_c$ manifests as bends in the linear IVC branch. Although undesirable, these weak links act as sensitive thermometers of junction temperature.

The impact of self-heating on SIS mixer performance depends strongly on LO power, which dominates over dc power and scales with frequency squared. Weak links can also perturb tuning by adding parasitic inductance, but unless operation frequency is very high or junction $J_c$ is low, this effect is modest. Overall, while self-heating is not catastrophic under typical operating conditions, careful substrate selection, process optimization, and thermal design remain essential for high-performance SIS mixers.



\section*{Acknowledgment}
The author would like to thank Yosuke Murayama and Takafumi Kojima for providing part of measurement samples and IVC data for comparison, and Hirotake Yamamori for useful discussion.

\ifCLASSOPTIONcaptionsoff
  \newpage
\fi



\bibliographystyle{IEEEtran}
%




\balance

\bibliography{SelfHeatingPaper}

%








\end{document}